\documentclass{iaus}
\usepackage{graphicx}

\title[The LCOGT Network] 
{The LCOGT Network}

\author[A. Shporer et al.]   
{A.~Shporer, T.~Brown, T.~Lister, R.~Street, \\ Y.~Tsapras, F.~Bianco, B.~Fulton, A.~Howell
}

\affiliation{Las Cumbres Observatory Global Telescope Network, 6740 Cortona Drive,\\ Suite 102, Santa Barbara, CA 93117, USA; {http://lcogt.net}} 

\pubyear{2011}
\volume{276}  
\pagerange{}
\jname{The Astrophysics of Planetary Systems}
\editors{A.~Sozzetti, M.~G.~Lattanzi \& A.~P.~Boss, eds.}
\begin{document}

\maketitle

\begin{abstract}
Motivated by the increasing need for observational resources for the study of time varying astronomy, the
Las Cumbres Observatory Global Telescope (LCOGT) is a private foundation, whose goal is to build a global network of robotic telescopes for scientific research and education. Once completed, the network will become a unique tool, capable of continuous monitoring from both the Northern and Southern Hemispheres. The network currently includes 2~$\times$~2.0~m telescopes, already making an impact in the field of exoplanet research. In the next few years they will be joined by at least 12~$\times$~1.0~m and 20~$\times$~0.4~m telescopes. The increasing amount of LCOGT observational resources in the coming years will be of great service to the astronomical community in general, and the exoplanet community in particular.
\end{abstract}


The LCOGT network currently consists of two 2.0~m telescopes: Faulkes Telescope South (FTS), located at Siding Spring Observatory, Australia, and Faulkes Telescope North (FTN), located on Mt.~Haleakala on the Hawaiian island of Maui. Two 0.4~m telescopes are also located within the FTN clamshell dome, and they are currently being commissioned.

In the next 1--2~years LCOGT will deploy telescopes at several sites, as shown in Figure~1. The deployment of the network in the Southern hemisphere will commence first. Construction has already begun at CTIO (Chile) and SAAO (South Africa). In each site  we plan to put 3~$\times$~1.0~m and 4--6~$\times$~0.4~m telescopes, starting 2011. The third node of our 0.4~m and 1.0~m network in the South will be in Australia, possibly at Siding Spring next to FTS, but we are considering other sites as well. In the North, we are now negotiating a site agreement with the IAC (Canary Islands, Spain) and McDonald Observatory (TX, USA). Our intention is to have an additional Northern node of the network in Asia, where a few possibilities are being investigated. 

\begin{figure}[b]
\begin{center}
 \includegraphics[width=4.6in]{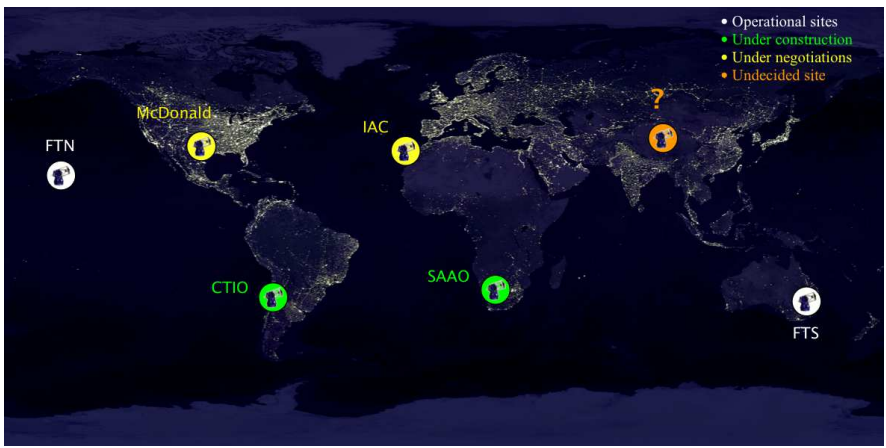} 
\caption{The planned LCOGT network, showing the location and status of the various sites.}
\label{fig1}
\end{center}
\end{figure}

Another site now being commissioned is the Byrne Observatory at UC's Sedgwick Reserve (BOS) in the Santa Ynez valley, approximately 30~miles from LCOGT's base in Goleta (CA, USA). This site currently has a 0.8~m telescope and will be used for testing new instruments and for education.

The LCOGT science team includes two UCSB faculty members and close to 10 postdocs and project scientists. Within the domain of time-variable astronomy LCOGT focuses primarily on two observational fields: Exoplanets and Supernovae. 
LCOGT is taking part in most of the important Supernovae surveys, including the Supernova Legacy Survey, Pan-STARRS, the Palomar Transient Factory, and the La Silla/QUEST Supernova search.
The study of exoplanets is carried out through observations of microlensing and transiting planets. 
Microlensing events are being monitored by the LCOGT-based network for the detection of microlensing planets, RoboNet (Tsapras et al.\ 2009).

LCOGT scientists are collaborating with most of the transiting planet surveys, including WASP, HATNet, TrES, MEarth, CoRoT and Kepler. A large fraction of FTN and FTS telescope time is devoted for observations of planetary transit candidates, in order to resolve candidates blended with nearby stars in survey images and obtain high precision photometry. This is usually done before or contemporaneously with the gathering of high resolution spectroscopic observations at other telescopes. Light curves obtained with LCOGT telescopes were part of the discovery of many of the currently known transiting planets, e.g., WASP-4b (Wilson et al.\ 2008), WASP-24b (Street et al.\ 2010), TrES-3 (O'Donovan et al.\ 2007), HAT-P-24b (Kipping et al.\ 2010) and CoRoT-9b (Deeg et al.\ 2010). LCOGT also participates in follow-up studies of known transiting exoplanets. For example, Hidas et al.\ (2010) and Shporer et al.\ (2010a) describe two follow-up campaigns led by LCOGT researchers where a complete coverage of HD~80606b 12 hour transits was obtained. 

In addition, members of the LCOGT science team are involved in the study of other variable stars (e.g., Lister et al.\ 2009, Shporer et al.\ 2010b) and open clusters (e.g., Cieza \& Baliber 2007; Fulton \& Baliber 2010, in preparation).

A few other projects are currently on-going at LCOGT. Among them is the commissioning of a high-speed camera mounted on FTN, to be used for lucky imaging and observations of Kuiper Belt Object occultations and other short time scale phenomena. We also note here that LCOGT is developing a medium-resolution (R=25,000) fiber-fed spectrograph, for stellar spectroscopy. It will reach a radial velocity accuracy on the order of 100~m/s, making it capable of identifying planetary transit false positives. The spectrograph is expected to be mounted on the Byrne Observatory telescope for on-sky testing during 2011, and will eventually be mounted on the 1.0~m telescopes.

\end{document}